\documentclass[journal]{IEEEtran}
\ifCLASSINFOpdf
\else
\fi
\usepackage{cite}
\usepackage{graphicx}
\usepackage[small]{caption}
\usepackage{subcaption}
\usepackage{array}
\usepackage{amsmath}
\usepackage{booktabs}

\begin{document}
%
\title{A 256kb 9T Near-Threshold SRAM With 1k Cells per Bit-Line and Enhanced Write and Read Operations}
\author{Ghasem~Pasandi,~\IEEEmembership{Student Member,~IEEE,}
       and Sied~Mehdi~Fakhraie
\thanks{Authors are with Nano-Electronics Center of Excellence, School
of Electrical and Computer Engineering, University of Tehran, Tehran,
I. R. Iran, 14395-515 e-mail: gh.pasandi, Fakhraie@ut.ac.ir.}
\thanks{Manuscript received June 23, 2014; revised October 13, 2014; accepted
November 11, 2014.} 
\thanks{Color versions of one or more of the figures in this paper are available
online at http://ieeexplore.ieee.org.
Digital Object Identifier 10.1109/TVLSI.2014.2377518}  

}

\markboth{IEEE Transactions on very large scale integration (VLSI) systems, Volume 23, Issue 11 (DOI: 10.1109/TVLSI.2014.2377518)}%
{Pasandi \MakeLowercase{\textit{et al.}}:A 9T Low-Voltage SRAM Cell with Enhanced Read and Write Operations}
%

\maketitle

\begin{abstract}
In this paper, we present a new 9T SRAM cell that has good write-ability and improves read stability at the same time. Simulation results show that the proposed design increases Read SNM (RSNM) and ${I_{on}}/{I_{off}}$ of read path by 219\% and 113\%, respectively at supply voltage of 300mV over conventional 6T SRAM cell in a 90nm CMOS technology. Proposed design lets us to reduce minimum operating voltage of SRAM ($VDD_{min}$) to 350mV, whereas conventional 6T SRAM cannot operate successfully with acceptable failure rate at supply voltages bellow 725mV. We also compared our design with three other SRAM cells from recent literature. To verify the proposed design, a 256kb SRAM is designed using new 9T and conventional 6T SRAM cells. Operating at their minimum possible VDDs, the proposed design decreases write and read power per operation by 92\%, and 93\%, respectively over the conventional rival. Area of proposed SRAM cell is increased by 83\% over conventional 6T one. However, due to large ${I_{on}}/{I_{off}}$ of read path for 9T cell, we are able to put 1k cells in each column of 256kb SRAM block, resulting in the possibility for sharing write and read circuitries of each column between more cells compared to conventional 6T. Thus, area overhead of 256kb SRAM based on new 9T cell is reduced to 37\% compared to 6T SRAM.
\end{abstract}

\IEEEpeerreviewmaketitle

\section{Introduction}

\IEEEPARstart{f}{or} MANY years, there was little interest in low power design and design trend was towards increasing the speed and working frequency of  digital systems \cite{sub-threshold}. Recently some applications such as implantable devices in man's body, portable applications and WSNs that need low-power circuits, increase the importance of low-power and ultra-low-power design \cite{K.Wang}.\\
SRAMs are an important part of most of the digital chips and consume a large percent of area and power of each chip \cite{Pasandi_CADS}, so decreasing the power and area of SRAMs can lead to decreasing the overall power and area of chips. Due to quadratic relation between power and supply voltage of transistors \cite{DIC_rabaey}, one effective and common method to reduce the power consumption is to decrease the supply voltage. By further decreasing the supply voltage, it will be lower than threshold voltage of transistors, so the circuit will operate in sub-threshold region. Unfortunately conventional designs cannot work properly in sub-threshold region, so new configurations are needed to let the digital systems work successfully in this region. In this paper, we present a new 9T SRAM cell that solves the problems of write and read operations at low supply voltages, thus let the SRAM to operate at smaller supply voltages.\\
The rest of this paper is organized as follows: In Section \ref{sec:problem} write and read operations problems of conventional 6T SRAM cell is mentioned and some recently published SRAM cells for resolving this problems are discussed. Section \ref{sec:prop} presents our new design for SRAM cell. In Section \ref{sec:sim}, simulation results for our design and also other SRAM cells are compared. Finally, Section \ref{sec:conc} concludes the manuscript.

\section{Problem statement and previously write and read enhanced SRAM cells}
\label{sec:problem}
Conventional 6T SRAM cell faces problems of write failures at low supply voltages \cite{Pasandi_ICEE}. Actually, at low supply voltages weak write-access transistors in this cell cannot overcome to the strong feedback of inverters of the cell. So, in duration of write operation, access transistors cannot enforce the input to the desired cell.\\
Also at low supply voltages, conventional 6T SRAM cell faces read failures during read operation, and content of the selected cell is subject to change with large likelihood. However, it is possible to increase the read stability of conventional 6T cell by using minimum size access transistors, but this will degrade write-ability of this cell. Moreover, problem of small Read Static Noise Margin (RSNM) is left un-resolved. Another challenge of conventional 6T SRAM cell is small ${I_{on}}/{I_{off}}$ of read-access transistors that doesn't let integrate large number of cells in each column of SRAM array. 
 At supply voltage of 200mV in a 0.13$\mu$m technology, this ratio decreases to 240 at the worst case pattern for data of cells in the same bit-line \cite{zhai_JSSC}. It means that at this supply voltage, number of cells sharing same bit-line must be enough smaller than 240 that the difference between ON current of selected cell and accumulated OFF currents of other cells can be distinguished by available sense amplifiers. It is desired to increase ${I_{on}}/{I_{off}}$ for read-access path to integrate more cells in the same bit-line. Integration of more SRAM cells in the same bit-line makes it possible to share write and read circuitries of each column between more cells, translating to saving area and power of these circuitries and hence total area and power of SRAM block.\\
To benefit from power lowering with supply voltage scaling, several designs were proposed in recently published papers. Some of them tried to resolve challenges related to write operation to improve the write-ability at small supply voltages. In general, there are two possible solutions to improve write-ability; the first one is to make write-access transistors stronger during write operation. Authors in \cite{sharifkhani_SVGND} used boosted (larger) voltage to control the write-access transistors during write operation, and authors in \cite{Tae-Kim_AVoltage} make access transistors stronger by employing reverse short channel effect (RSCE) (using access transistors with larger (3X) channel length). The second solution to improve write-ability of conventional SRAM cell is to make the feedback loops between inverters of the cell weaker or to brake this feedback loop during write operation \cite{Aly_SOC,Pasandi_ICEE2014_1,zhai_JSSC,Kouichi_JSSC,Masanao_JSSC}. \\
Some other designs tried to resolve problems related to read operation. To decrease the read failure probability, a 7T \cite{Takeda_JSSC} and two 8T SRAM cells were proposed \cite{saeidi2014subthreshold,Hassanzadeh_DTIS}. To increase Read SNM (RSNM), several forms of buffering read (sensing voltage of internal node by gate of a transistor) were proposed \cite{9T_Liu,simplest8T_chang,LP10T_Hasan,Takeda_JSSC}. 
In these cells due to buffering read operation, RSNM is increased to the level of Hold SNM (HSNM), but in all of these designs, problem of small ${I_{on}}/{I_{off}}$ for read-access path is still remained un-resolved.\\
Authors in \cite{calhoun_JSSCC_256k, Kim_10T, Verma_256kb} used a modified version of buffering read that improves ${I_{on}}/{I_{off}}$ of read access path and RSNM at the same time. 
\section{Our Proposed Design}
\label{sec:prop}
\begin{figure}[!t]
\centering
\includegraphics[width=0.5\textwidth]{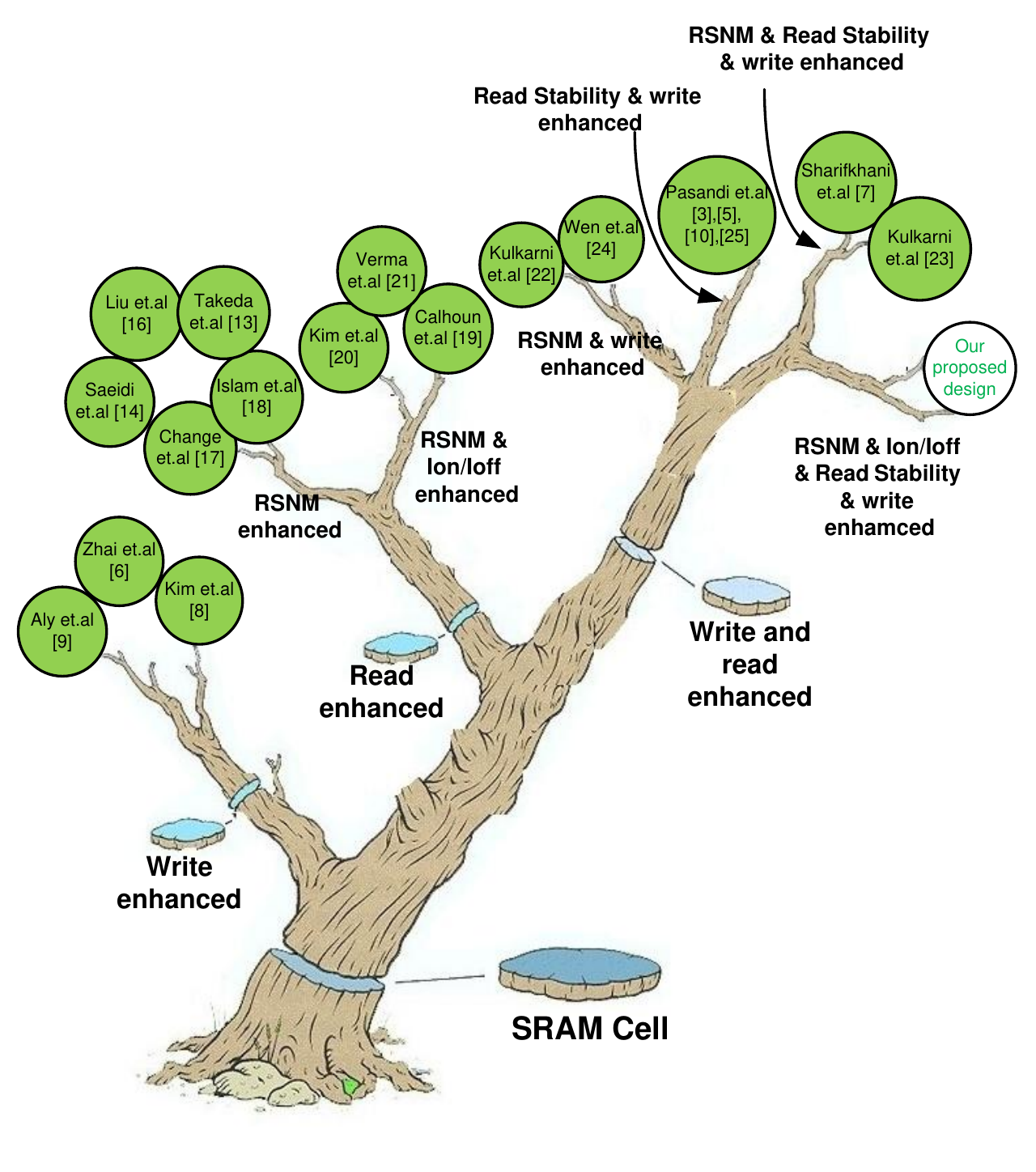}
\caption{Tree diagram for different SRAM cells based on the improvement they suggest.}
\label{fig1}
\end{figure}

\begin{figure*}
        \centering 
        \begin{subfigure}[!t]{0.33\textwidth}
                \centering
                \includegraphics[width=\textwidth]{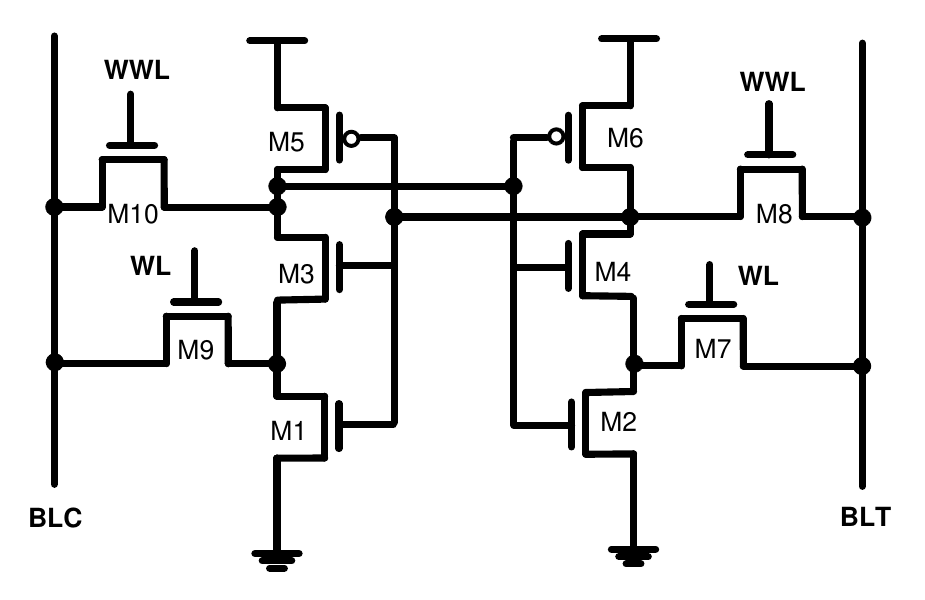}
                \caption{New schmitt-trigger based 10T SRAM cell (ST-2)\cite{ST-2}}
                \label{fig2-1}
        \end{subfigure}
        \begin{subfigure}[!t]{0.32\textwidth}
                \centering
                \includegraphics[width=\textwidth]{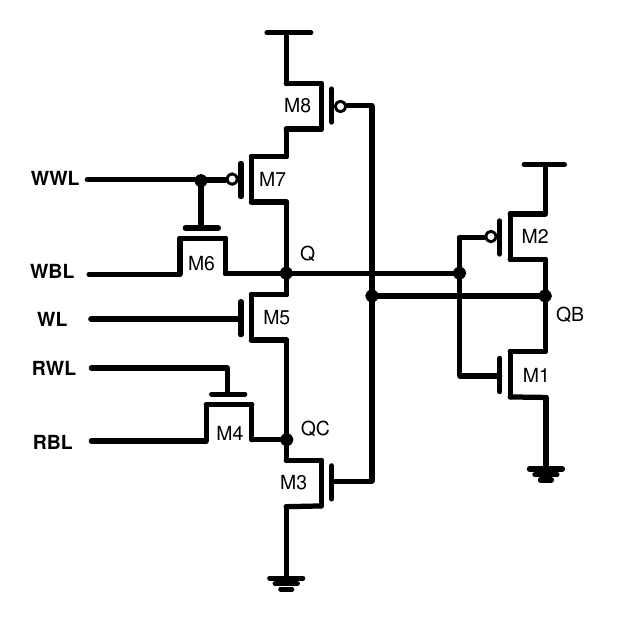}
                \caption{Single-ended 8T SRAM cell (WEN cell)\cite{Wen}}
                \label{WEN_cell}
        \end{subfigure}
        \begin{subfigure}[!t]{0.33\textwidth}
                \centering
                \includegraphics[width=\textwidth]{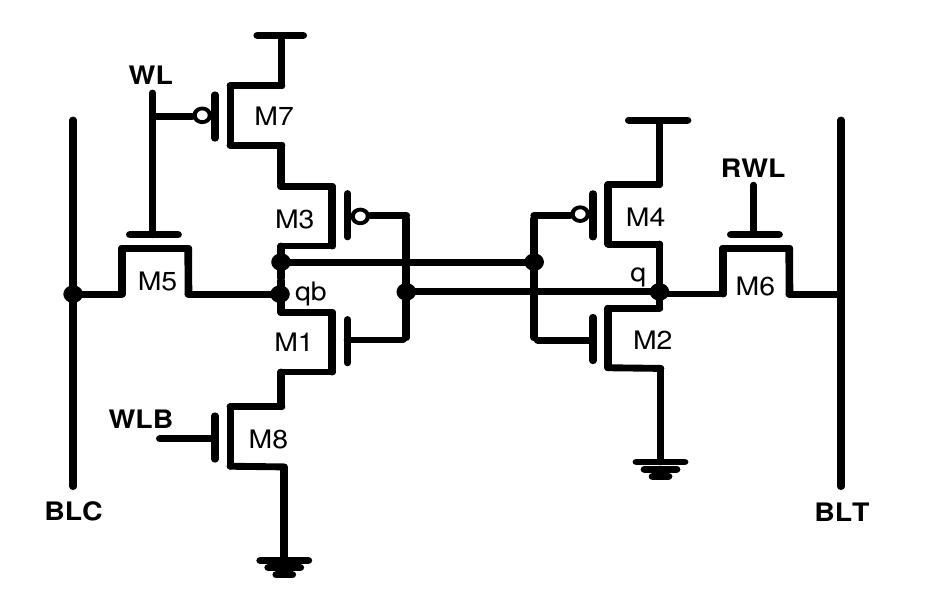}
                \caption{Write and Read Enhanced 8T SRAM cell (WRE8T) \cite{Pasandi_TED}}
                \label{WRE8T_cell}
        \end{subfigure}%
        \caption{Three write and read enhanced SRAM cells from recent literature used for comparison with this work.}\label{WR_enhanced}
\end{figure*}

In previous section, some techniques used in several SRAM cells were introduced. Each of these cells can improve write or read operations at lower supply voltages. There are some SRAM cells proposed in recently published papers that can improve write and read operations at the same time. Kulkarni and Roy changed the design in \cite{ST10T_JSSC}, and proposed a modified version of schmitt-trigger based SRAM cell (ST-2 hear after)\cite{ST-2}. This SRAM cell increases write-ability due to stacked transistors in the inverters of the cell and improves read operation by using individual added access transistors. Wen et al. \cite{Wen} introduced single-ended 8T SRAM cell (WEN cell hear after) that can improve write and read operations at the same time. We proposed an 8T SRAM cell in our previous work \cite{Pasandi_TED} that improves write-ability due to ability of the cell to make one of the inverters of the cell weaker during write operation (WRE8T hear after). This cell also improves read operation due to single ended write and read operations that let to avoid access transistors sizing conflict. Fig. \ref{fig1} shows different designs for SRAM cell based on the improvements that they suggest. Fig. \ref{WR_enhanced} shows the circuit diagram of previously write and read enhanced SRAM cells.\\
In this paper, we present a new 9T SRAM cell that improves write and read operations at the same time. In the next sub-sections we will explain the write and read enhanced techniques used in our proposed Write and Read Enhanced 9T SRAM cell (called WRE9T).
\subsection{Write Enhancement Techniques in Proposed SRAM Cell}
\label{sec:prop:a}
\begin{figure}[t]
\centering
\includegraphics[width=0.47\textwidth]{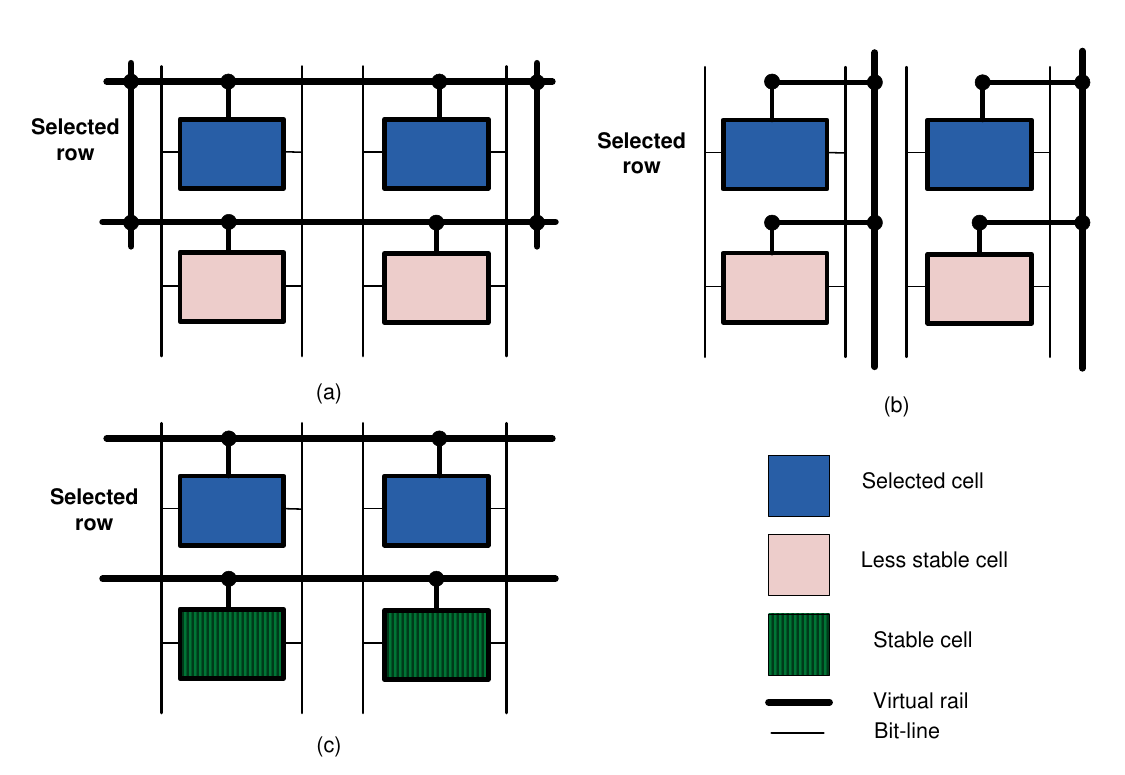}
\caption{Three possible architectures for sharing power and ground rails (a) sharing virtual rails for all cells in the array. Each column (b) and each row (c) has individual virtual rail and all cells in the same column (row) share it.}
\label{possible_integration}
\end{figure}

\begin{figure}[t]
\centering
\includegraphics[width=0.4\textwidth]{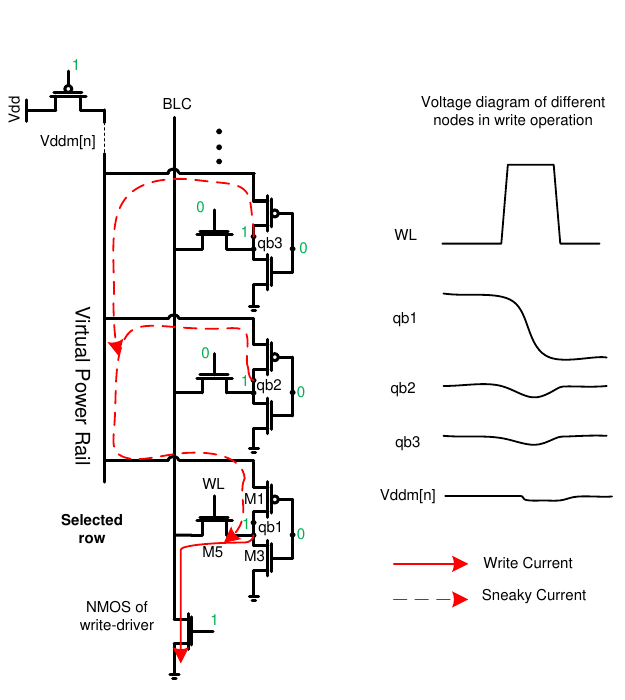}
\caption{Sneaky current in write operation for architecture of SRAM used in \cite{Masanao_JSSC} that can change voltage of internal nodes in half selected cells.}
\label{sneaky_current_2}
\end{figure}

In the SRAM cell used in \cite{Pasandi_TED}, one NMOS and one PMOS transistor were added to each cell of SRAM that become OFF during write operation and provide floating power and ground rails. In this cell, to control the gate of added NMOS transistor, a separate word-line is needed. These added transistors and word-line impose area overheads. The point is that, it is not necessary to add these two transistors to each cell to have floating power and ground rails. We can use one NMOS and one PMOS transistor that provide these floating rails and distribute these rails among the sharing cells. There are some possible strategies for sharing the floating rails. Fig. \ref{possible_integration} shows three architectures. In parts (a) and (b), there are cells that are not selected but become weaker during write operation. The less stable cells are shown with light color in this figure and are called half selected cells (their WL are 0 but power rail of them are floated). Simulations at supply voltage of 500mV show that mean of SNM of half selected cells in 1000 runs of Monte Carlo simulations is 0.33mV while it is 167.4mV for not-selected cells. This margin is very small and even at room temperature, thermal noise can flip the content of the cell.\\
Authors in \cite{Masanao_JSSC} used architecture of Fig. \ref{possible_integration} Part(b) for designing the SRAM and shared power line of all cells in the same column. In this design there is a possible state for content of the cells in the same column, that will lead to increased write power due to existence of sneaky current. This strategy and sneaky current are shown in Fig. \ref{sneaky_current_2}. In this figure, only the left inverter of the SRAM cells are shown. As seen, to write '0' to the selected cell, if voltage of right internal nodes in the other cells sharing the column are set to '0', the PMOS transistors in these cells will be ON and let the sneaky current to pass through shared floating line and discharge the voltage of internal nodes in not-selected cells. Thus, activity of nodes and hence, write power will increase. To test the effect of this sneaky current on power consumption, a column of SRAM cells with one selected and 64 other non-selected cells were designed using this architecture and write power for writing '0' was extracted. Two strategies were considered. Case 1: voltage of right internal nodes in non-selected cells are '0' so sneaky current will exist, and case 2: voltage of right internal nodes in non-selected cells are 'Vdd' so, path for sneaky current will not exist. Extracted power for case 1  is $1.18\mu$W and for case 2 is 0.1$\mu$W.\\

Due to discussed problems for architectures of Fig. \ref{possible_integration} Parts (a) and (b), we choose third architecture (Fig. \ref{possible_integration} part (c))for implementing our SRAM. Even in this architecture, state of half selection will exist if we decide to select each bit individually. To solve the disturbance in this state, we apply our half selection disturb-free scheme introduced in \cite{Pasandi_TED}. In this scheme, before first write operation in consecutive writes, the content of each cell in the selected row are read and then, using embedded inverters in each column, they are written back on the corresponding write bit-lines.  
\subsection{Read Enhancement Techniques in Proposed SRAM Cell}
\label{sec:prop:b}
\begin{figure}[!t]
\centering
\includegraphics[width=0.4\textwidth]{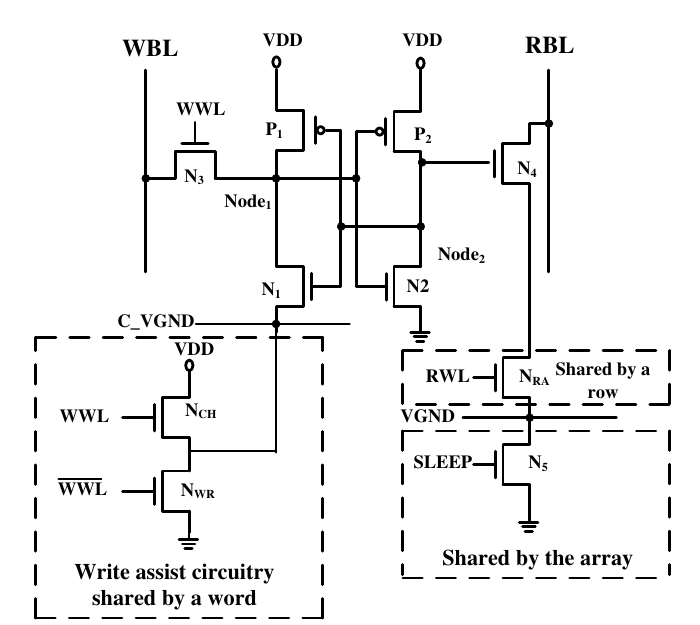}
\caption{Power gated 6T SRAM cell presented in \cite{Hailong}.}
\label{Hailong_cell}
\end{figure}

\begin{figure}[!t]
\centering
\includegraphics[width=0.5\textwidth]{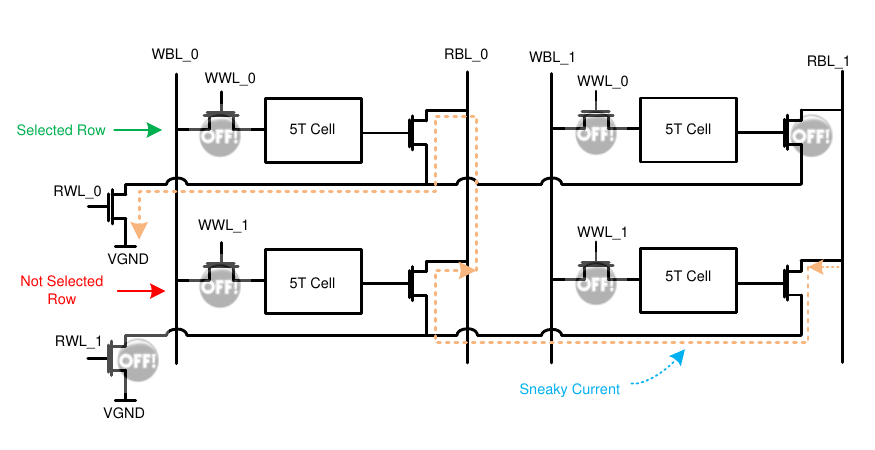}
\caption{A possible state that can discharge RBL\_1 bit-line falsely (another type of sneaky current).}
\label{sneaky_current}
\end{figure}

One method to increase RSNM is sensing voltage of internal nodes of SRAM cell through gate of a transistor in read operation. The simplest way of buffering is   using two transistors that one of them (buffer transistor) senses internal voltage of SRAM cell, and the other one is controlled by RWL signal \cite{simplest8T_chang}. In the case that buffer transistor is ON, for un-selected cell, leakage current is large. In \cite{Hailong,9T_verlika} three transistors are used in read path. Depending on the internal voltage of SRAM cell, there are two or three OFF transistors in read path for un-selected SRAM cells. Equation (\ref{eq-Isub}) shows the dominating part of leakage current, that is subtreshold current \cite{LowPower_Rabaey}. For $V_{DS}$ larger than 50mV, we can ignore the second term to reach to Equation (\ref{eq-Isub2}) \cite{LowPower_Rabaey}. 

\begin{equation}
I_{Dsub} = I_s10^{\frac{V_{GS}-V_T+\lambda V_{DS}}{S}}(1-10^{-\frac{\eta V_{DS}}{S}})
\label{eq-Isub}
\end{equation}
\begin{equation}
I_{Dsub} = I_s10^{\frac{V_{GS}-V_T+\lambda V_{DS}}{S}}
\label{eq-Isub2}
\end{equation}
Now, the worst case leakage current for two and three stacked transistors (one of stacked transistors is ON) can be expressed by Equations (\ref{eq-Isub3}), and (\ref{eq-Isub4}), respectively. Assuming that $V_T=300mV$, $\lambda=1.5$, and $S=80mV$ (typical values \cite{LowPower_Rabaey}), for VDD=1.0V, ratio of the worst case leakage current for three stacked transistors ($I_{31}$), to the worst case of two stacked transistors ($I_{22}$) can be expressed by $I_{31}/I_{22}=10^{\frac{-\lambda^4-3\lambda^3-2\lambda^2}{6\lambda^3+9\lambda^2+5\lambda+1}VDD/S}=10^{-5}$. As seen, leakage current is decreased by exponential order if one stacked transistor is added to read path.\\
 In \cite{Hailong}, one of three stacked transistors in the read path is shared among all cells in the array, so it is OFF only in idle modes, so, the worst case for leakage current flowing through read path is that two of stacked transistors are ON and one of them is OFF. Leakage current for this case can be expressed by Equation (\ref{eq-Isub5}), that shows 80\% increasing compared with Equation (\ref{eq-Isub4}).
\begin{equation}
I_{22} = I_s10^{\frac{-V_T+\frac{\lambda^2+\lambda+1}{2\lambda+1} V_{DD}}{S}}
\label{eq-Isub3}
\end{equation}
\begin{equation}
I_{31} = I_s10^{\frac{-V_T+\frac{\lambda^3+\lambda^2+2\lambda+1}{3\lambda^2+3\lambda+1} V_{DD}}{S}}
\label{eq-Isub4}
\end{equation}
\begin{equation}
I_{32} = I_s10^{\frac{-V_T+\frac{\lambda^3+2\lambda^2+\lambda}{3\lambda^2+3\lambda+1} V_{DD}}{S}}
\label{eq-Isub5}
\end{equation}
In addition to this increment of leakage current in SRAM cell of \cite{Hailong}, there is a possible state that read bit-line will be discharged falsely. Fig. \ref{Hailong_cell} shows SRAM cells used in \cite{Hailong}. In Fig. \ref{sneaky_current}, a possible state is illustrated that can lead to falsely discharging the bit-lines and cause read fault. In this figure, RBL\_1 bit-line must not be discharged because stored voltage in the second cell in the first row is 0; but the shown sneaky current can discharge this bit-line and leads to read fault.\\
\begin{figure}[!b]
\centering
\includegraphics[width=0.45\textwidth]{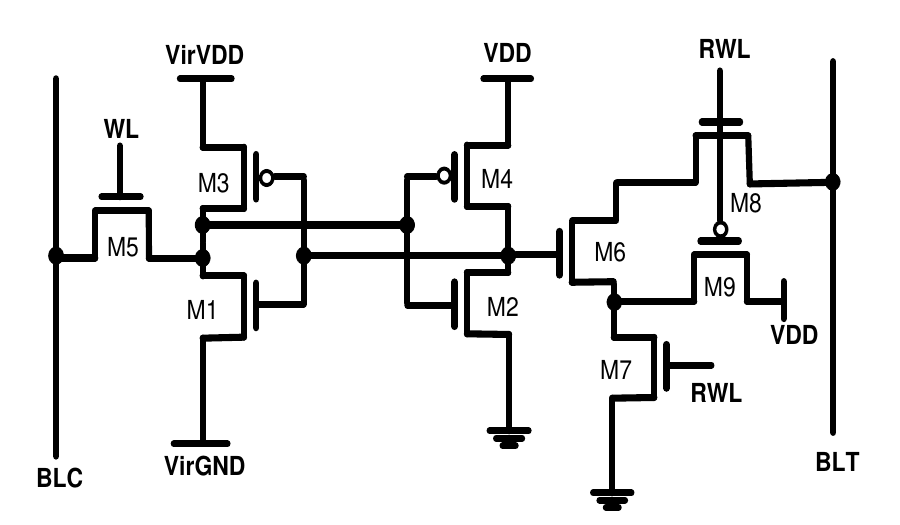}
\caption{Proposed write and read enhanced 9T SRAM cell (WRE9T).}
\label{WRE9T}
\end{figure}
In SRAM cell of \cite{9T_verlika}, the above sneaky current does not exist, and for not selected cells, there are two or three OFF transistors in the read path, so leakage current for not selected cells is small. However, leakage current for a selected SRAM cell that are storing a '0' on the right internal node is similar to the case that was expressed by Equation (\ref{eq-Isub5}); because in this case, there are two ON and one OFF transistors in the read path. So, in this case large leakage current can falsely discharge the read bit-line.\\ 
To solve the challenges of SRAM cells used in \cite{Hailong} and \cite{9T_verlika}, we are proposing a new design for SRAM cell. Fig. \ref{WRE9T} shows circuit diagram of our 9T SRAM cell. Fig. \ref{my_readpath} shows the read path of our design. In this design, added PMOS transistor M9 is ON for not selected cells so the voltage of node Qc will rise to reach VDD for these cells. This will reduce the leakage current passing through M6 for two reasons. The first reason is that by increasing the voltage of source of M6, its threshold voltage will increase according to Equation (\ref{eq-Vth}) \cite{DIC_rabaey}. Increasing $V_{th}$ will decrease the sub-threshold leakage current $I_{sub}$ (that is the main component of leakage current) due to Equation (\ref{eq-Isub}). The second reason behind decreasing leakage current is that by increasing voltage of source of M6, $V_{GS}$ of this transistor will decrease and due to Equation (\ref{eq-Isub}) this will reduce sub-threshold current $I_{sub}$ exponentially.\\
As mentioned earlier, challenge of design used in \cite{9T_verlika} is that large leakage current flows through read access path for an specific data pattern in selected cells. In this case, despite the stored data in the selected cell, leakage current can discharge the read-bit-line capacitance falsely. In our design, for selected cells, added PMOS transistor M9 is OFF but still passes some leakage current to node Qc. This will increase voltage of node Qc a little, and due to Equations (\ref{eq-Isub}) and (\ref{eq-Vth}), reduce leakage current flowing through M6.
\begin{equation}
V_{T} = V_{T0} + \gamma (\sqrt{|-2\phi_F + V_{SB}|} - \sqrt{|-2\phi_F|})
\label{eq-Vth}
\end{equation}
\\Fig. \ref{IonOFF_WVC} shows ON to OFF current ratios for read buffered transistors in read path for the three designs (Typical Corner, and T=27 $^\circ$C). In this figure, voltage of internal node is assumed to be VDD. As seen, ${I_{on}}/{I_{off}}$ of read path for our proposed design is considerably larger than the two others. For example, at VDD=0.8V, ${I_{on}}/{I_{off}}$ of WRE9T design is 8.66 times larger than 9T cell in \cite{9T_verlika}. Thus, it is possible to integrate larger number of cells sharing same bit-line in our design compared to two others, resulting in saving the area and power of peripheral circuits used for each column in SRAM array.
\begin{figure*}[!t]
        \centering
        \begin{subfigure}{0.23\textwidth}
                \centering
                \includegraphics[width=\textwidth]{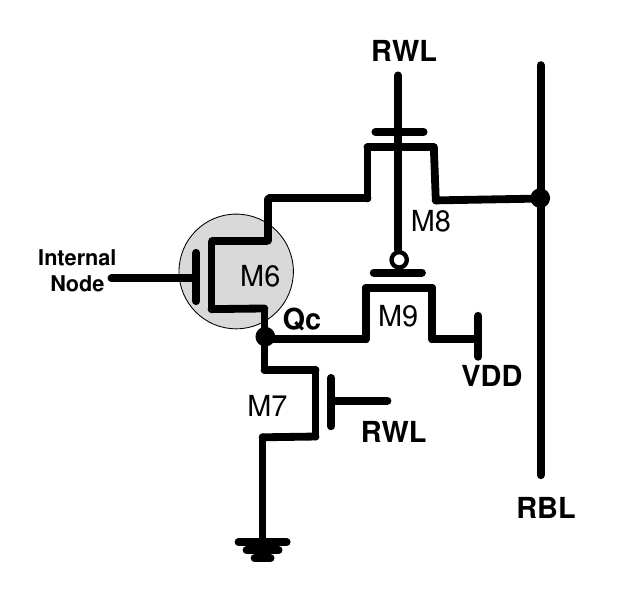}
                \caption{WRE9T read path.}
                \label{my_readpath}
        \end{subfigure}%
        \begin{subfigure}{0.17\textwidth}
                \centering
                \includegraphics[width=\textwidth]{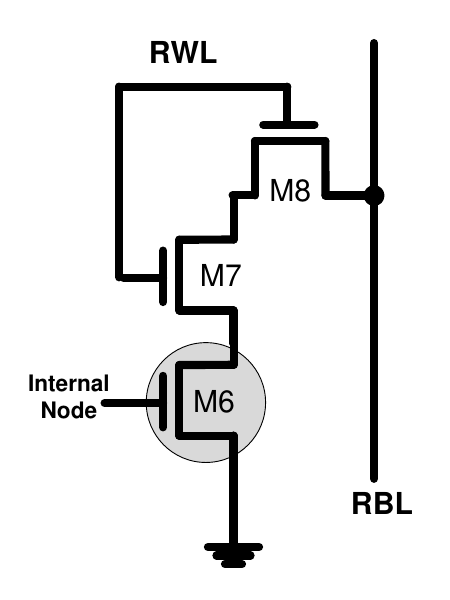}
                \caption{Read path of the cell in \cite{9T_verlika} (Verlika).}
                \label{verlika_readpath}
        \end{subfigure}
        \begin{subfigure}{0.18\textwidth}
                \centering
                \includegraphics[width=\textwidth]{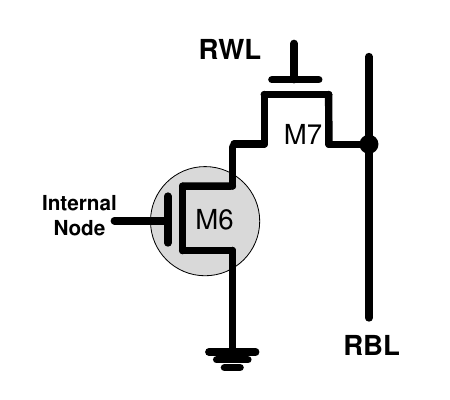}
                \caption{Read path of the cell in \cite{simplest8T_chang} (Chang)}
                \label{chang_readpath}
        \end{subfigure}
        \begin{subfigure}{0.4\textwidth}
                \centering
                \includegraphics[width=\textwidth]{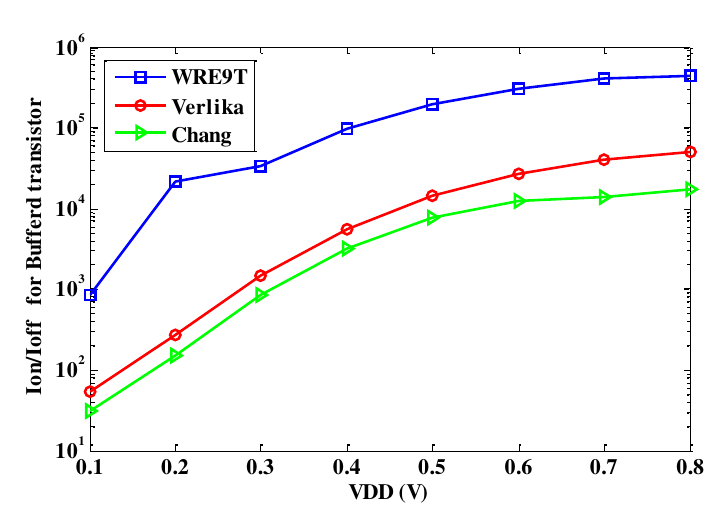}
                \caption{${I_{on}}/{I_{off}}$versus VDD for buffered transistor in read path for three designs.}
                \label{IonOFF_WVC}
        \end{subfigure}
        \caption{Comparing ${I_{on}}/{I_{off}}$ of read bufferd transistor for three different designs.}\label{IonOFF_Circuit_graph}
\end{figure*}
\section{Simulation Results and Discussion}
\label{sec:sim}
Since our proposed SRAM cell in this paper is write and read enhanced, thus in this part we extract and compare different working parameters of this cell with other write and read enhanced SRAM cells and also conventional 6T SRAM cell. HSPICE 2011 was used for simulations and circuits were designed in a 90nm industrial CMOS technology.\\
To make the comparison of conventional 6T SRAM cell with our proposed SRAM cells more fair, we also consider Iso-Area 6T (I.A.6T) SRAM cell \cite{ST-2}. In I.A.6T  SRAM cell, transistors were up-sized so that the area of this cell becomes equal to WRE9T SRAM cell.
\subsection{${I_{on}}/{I_{off}}$ of read path}
\label{sim:IonOFF}
\begin{figure}[!t]
\centering
\includegraphics[width=0.4\textwidth]{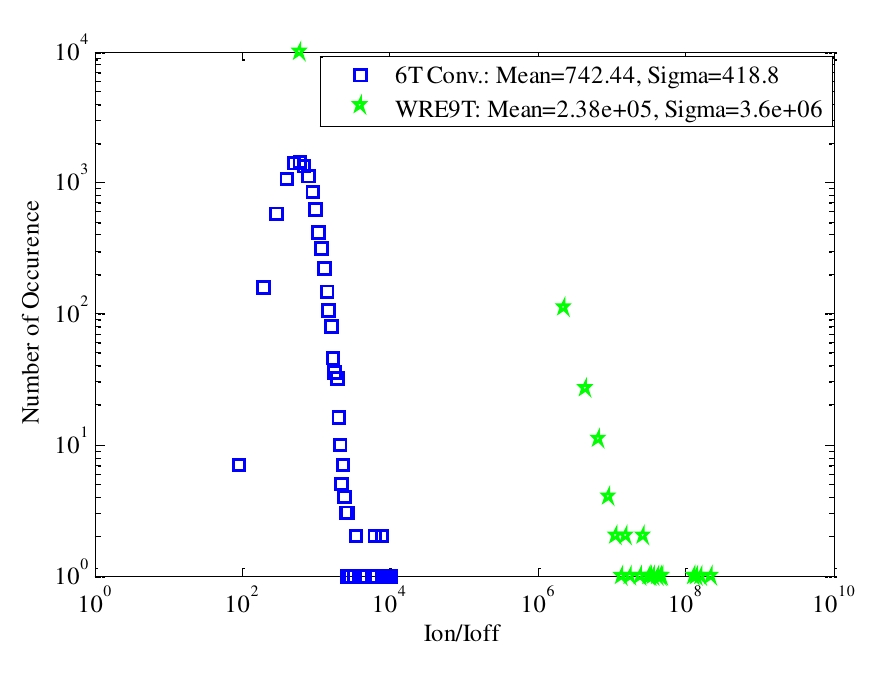}
\caption{Distribution of ${I_{on}}/{I_{off}}$  for read path of WRE9T and conventional 6T SRAM cells.}
\label{IonOFF_1}
\end{figure}

\begin{figure}[!t]
\centering
\includegraphics[width=0.43\textwidth]{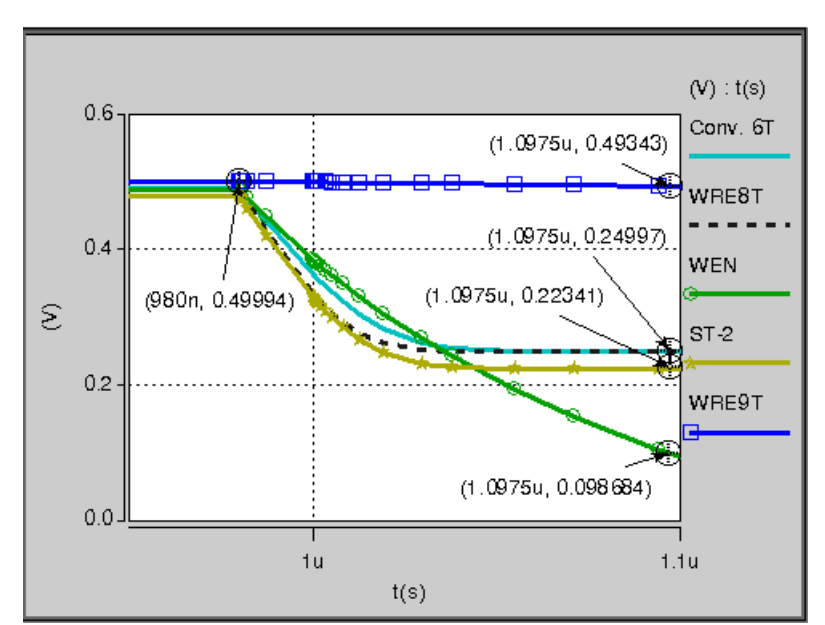}
\caption{Voltage of read bit-line (that should not be discharged according to stored voltage in the selected cell).}
\label{R-bit-line-voltage}
\end{figure}

\begin{figure}[!t]
\centering
\includegraphics[width=0.43\textwidth]{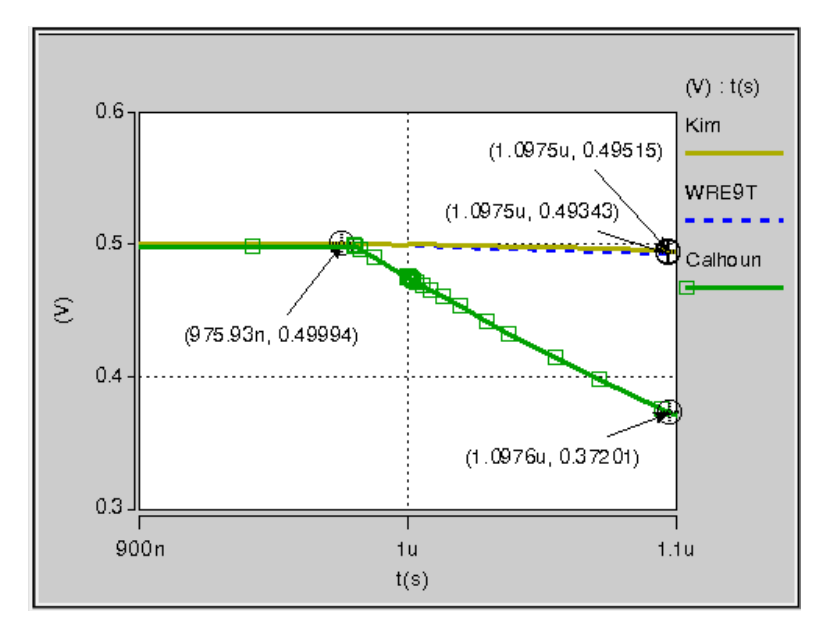}
\caption{Voltage of read bit-line for WRE9T, Calhoun et al. \cite{calhoun_JSSCC_256k} and Kim et al. \cite{Kim_10T}.}
\label{R-bit-line-voltage_2}
\end{figure}

Fig. \ref{IonOFF_1} shows distribution of ${I_{on}}/{I_{off}}$ for proposed SRAM cell and conventional one in presence of process variation at supply voltage of 300mV. As seen, mean of ${I_{on}}/{I_{off}}$ for proposed cell is larger than conv. 6T. Worst case ratio for proposed design is 1080 and is 78 for 6T. 
\\For the case that voltage of storage node is such that the related bit-line should not be discharged, leakage current of selected and not selected cells become very important. As the leakage of read path become larger, voltage of read bit-line decreases much faster and this will lead to read the stored data falsely. Fig. \ref{R-bit-line-voltage} shows voltage of read bit-line (for differential cells, voltage of the bit-line that should not be discharged) when it should not be discharged at supply voltage of 500mV. As seen, voltage drop for WRE9T cell since read word-line asserted ($@$ 980 ns) until 120 ns later, is almost 7mV, whereas for the best of others (conventional 6T), this voltage drop is almost 250mV. Voltage of read bit-line for ST-2 and WRE8T reach to VDD/2 after 30ns. Thus, if voltage of this bit-lines are read after this time, read error will occur. In our design, by using stacking effect and also by adding the PMOS transistor M9, leakage current through read path is very small. In the read path of SRAM cells used in \cite{calhoun_JSSCC_256k} and \cite{Kim_10T} also the path between read bit-line and ground is broken, so it is expected that voltage drop of read bit-line for these cells becomes small similar to our design. Fig. \ref{R-bit-line-voltage_2} shows the voltage of read bit-line for WRE9T cell and those of \cite{calhoun_JSSCC_256k} and \cite{Kim_10T}. Voltage drop for Calhoun et al. cell \cite{calhoun_JSSCC_256k} after 120ns is almost 127mV and remains above VDD/2. Voltage drop of Kim et al. cell \cite{Kim_10T} is almost equal to WRE9T cell and is acceptable.
\subsection{Read, Write, and Hold SNMs}
\label{sec:SNM}
\begin{figure}[!t]
\centering
\includegraphics[width=0.4\textwidth]{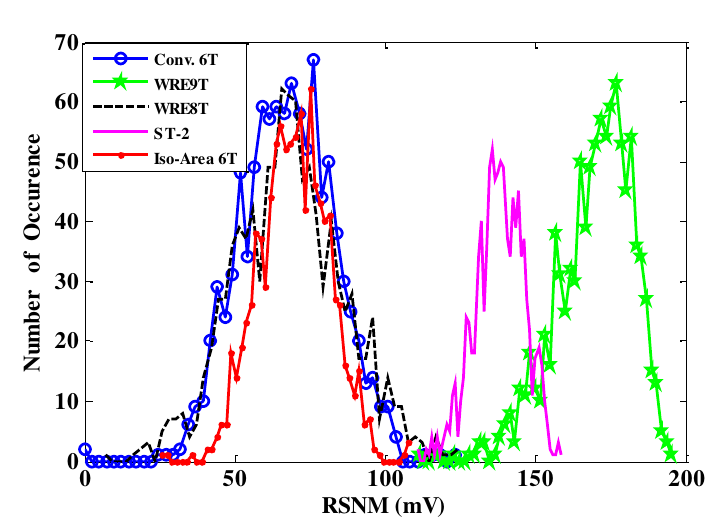}
\caption{Read SNM distribution of WRE9T and 4 other SRAM cells at supply voltage of 500mV (Monte Carlo simulations were performed for modelling fabrication process variation effect.)}
\label{RSNM_distribution_2}
\end{figure}

\begin{figure}[!t]
\centering
\includegraphics[width=0.4\textwidth]{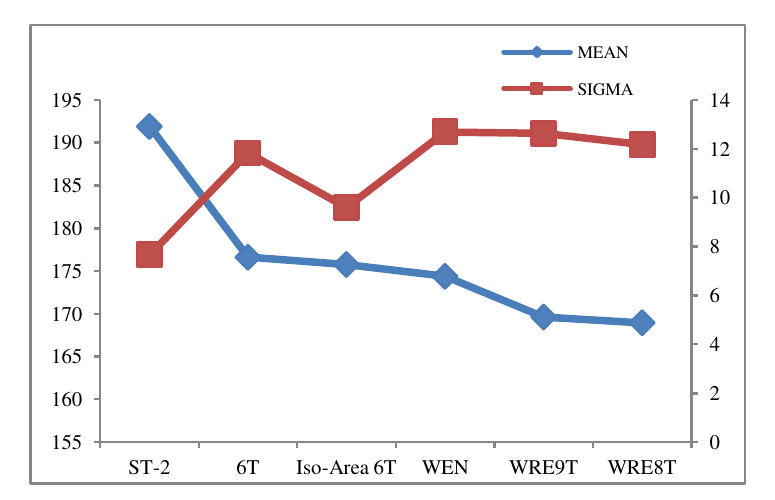}
\caption{Mean and sigma of HSNM (mV) for different  SRAM cells in 1000 runs of Monte Carlo simulations.}
\label{HSNM_Mean_sigma}
\end{figure}

\begin{figure}[!t]
\centering
\includegraphics[width=0.4\textwidth]{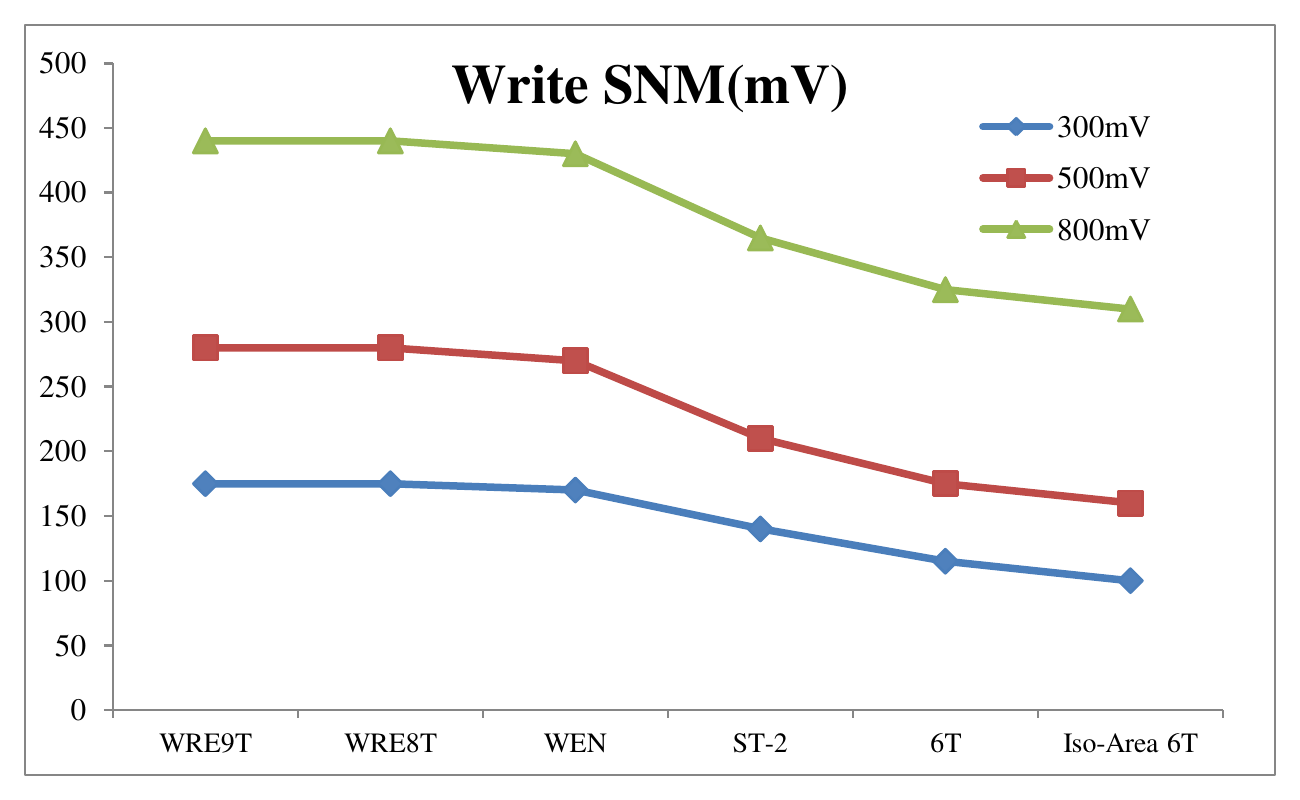}
\caption{Write noise margin for different designs at three supply voltages.}
\label{WNM_fig}
\end{figure}

One of the most important drawbacks of conventional 6T SRAM cell is small read SNM (RSNM) at low supply voltages. This does not let scaling of supply voltage and going to sub-threshold region to benefit from power saving of it \cite{Gupta_Digital_Computation}. Fig. \ref{RSNM_distribution_2} shows distribution of RSNM for WRE9T and 4 other SRAM cells at supply voltage of 500mV. Mean of RSNM for our design is better than conventional and ST-2 by 2.53X and 1.23X, respectively. In WEN SRAM cell \cite{Wen}, voltage of left node in the cell is kept floating during read operation. At typical corner and room temperature, leakage of PMOS pull up transistors in this cell is considerable that leads to destroying the dynamic voltage of left storage node. Thus, RSNM for this cell is very small at these conditions (Mean is 0.3mV and sigma is 1.8mV).
\\Fig. \ref{WNM_fig} shows write noise margin (WNM) of compared cells at three different supply voltages. WNM is a metric to measure the write-ability of SRAM cells. Having larger WNM for a cell means that write-ability is larger for the cell. In WRE9T, WRE8T and WEN SRAM cells, feedback loops in the cells  are interrupted during write operation that results in more successful write operations for these cells. This is the reason of larger WNM for these three SRAM cells in Fig. \ref{WNM_fig}.
\\Hold SNM (HSNM) is not such challenging and is normally large enough at low supply voltages. Fig. \ref{HSNM_Mean_sigma} shows mean and sigma HSNM for different SRAM cells at supply voltage of 500mV. Except ST-2 cell that have better HSNM compared to others, HSNMs of others are not far from each other, and the difference between mean of the best design and the worst one is about 7mV.
\subsection{Minimum Operating Voltage; $VDD_{min}$}
\label{sec:VDDmin}
To benefit from power saving of voltage scaling, it is desired to design SRAMs that can operate successfully at lower supply voltages. There are two different methods to determine minimum supply voltage for SRAMs. The first one is based on SNM \cite{ST-2} that does not capture the transient behaviour of the cell. The second one considers the transient behaviour of the cell but uses some approximations that lead to large errors at the tail of the distribution where the sensitive failure probabilities exist \cite{Vmin_Statistical}. In this paper, minimum operating voltage is extracted while the circuits operate at their real operating modes so that transient behaviour is captured. Assuming not having error correcting code (ECC) in data, and considering the size of SRAM to be 1Mb, the failure probability is considered $10^{-6}$  \cite{Vmin_Statistical}.\\
To have accurate results and to model the process variation effect, Monte Carlo (MC) simulations were used. In this simulations different parameters such as threshold voltage, channel length, gate oxide thickness, carrier mobility, and channel doping concentration were varied with Gaussian distributions.\\
To extract the failure probability in above simulations, dynamic criteria proposed in \cite{Khalil_VDDmin} was used and the outputs were measured at the end of the operation cycle. Table \ref{VDDmin_table} shows minimum operating voltages for different modes of operation. $VDD_{min}$ for each design is the maximum of these minimum voltages. As seen, $VDD_{min}$ for our proposed design, WRE9T is lower than others.
\begin{table}[b]
  \centering
  \caption{Minimum supply voltage for write, read, access, and data retention and $VDD_{min}$ (V).}
  \resizebox{\linewidth}{!}{
    \begin{tabular}{c c c c c c c}
    \toprule
    Design & WRE9T & WRE8T & 6T   & ST-2  & WEN & I.A.6T\\
    \midrule
    Read  & 0.320  & 0.430  & 0.725 & 0.340  & 0.355 & 0.390 \\
    Access & 0.310  & 0.405 & 0.555 &   1.20$^\star$  & 1.20$^\star$  & 0.375 \\
    Write & 0.340  & 0.320  & 0.455 & 0.375 & 0.315 & 0.480 \\
    Retention & 0.350  & 0.350  & 0.350  & 0.375 & 0.350 & 0.275 \\
    \hline
    $VDD_{min}$ & 0.350  & 0.430  & 0.725  &   1.20    & 1.20  & 0.480\\
    \bottomrule
    \end{tabular}}%
    \space\space\space\space\space$^\star$These cells could not reach to failure probability of $10^{-6}$ at lower voltages, thus, nominal voltage was chosen for them.
  \label{VDDmin_table}%
\end{table}%
\subsection{Power Consumption}
\label{sec:power}
\subsubsection{Leakage Power}
\begin{figure}[!t]
\centering
\includegraphics[width=0.4\textwidth]{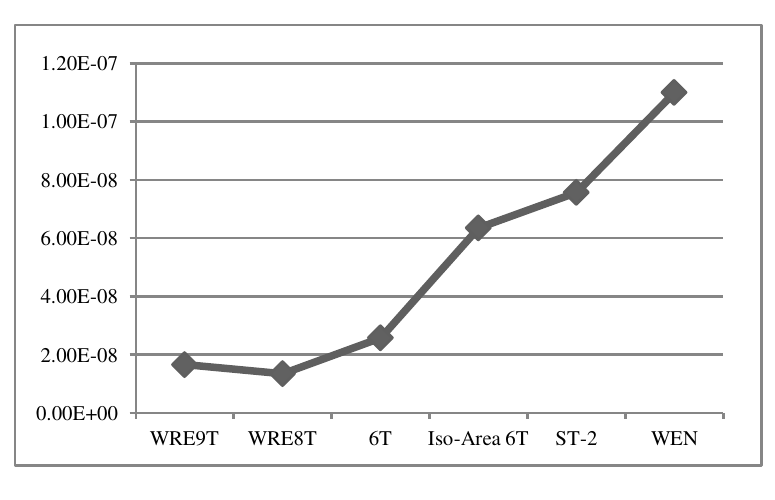}
\caption{Leakage power consumption (W) of different SRAM cells at T= 110$^\circ$C.}
\label{leakage_power_fig}
\end{figure}

In Section \ref{sec:VDDmin}, minimum operating voltage of different SRAM cells were extracted. Assuming that the mentioned SRAMs are used in low-power circuits, they are expected to operate at their $VDD_{min}$; thus, we extracted leakage power of each cell at its $VDD_{min}$. Fig. \ref{leakage_power_fig} shows the leakage power of different cells. These results were extracted in typical corner and at temperature of 110$^\circ$C.
\subsubsection{Total Power Consumption for Single Write Operation}
\begin{figure}[!t]
\centering
\includegraphics[width=0.4\textwidth]{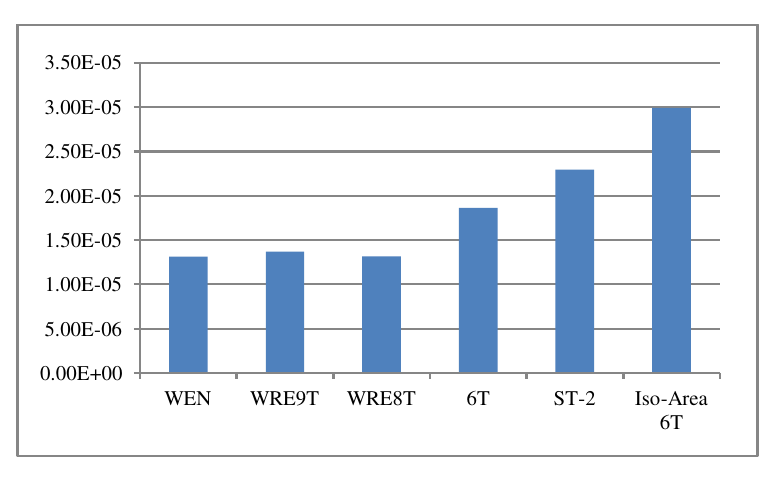}
\caption{Comparing total power consumption (W) of different SRAM cells for single write operation at VDD=500mV.}
\label{write_power_fig}
\end{figure}

Fig. \ref{write_power_fig} shows power consumption for single write operation for different SRAM cells. This power consumption is average of power consumption for writing one '0' and one '1' to the desired SRAM cell. As seen, power consumption for WRE8T, WRE9T, and WEN SRAM cells are smaller than three others. This is mainly due to single ended write operation in these SRAM cells, that leads to have smaller charging-discharging on write-bit-lines (for single ended cells there is one write-bit-line compared with differential ones that there are two write-bit-lines).
\begin{table}[t]
  \centering
  \caption{Single cell area comparison for different SRAM cells in 90nm CMOS technology.}
  \resizebox{\linewidth}{!}{
    \begin{tabular}{c c c c c c c}
    \toprule
    Design & WRE9T & WRE8T & 6T   & ST-2  & WEN & I.A.6T\\
    \midrule
    Area($\mu$m$^2$) & 3.72  &  2.89  &  2.03  & 3.99  & 3.45  & 3.72\\
    \bottomrule
    \end{tabular}}%
  \label{Layout_Areas}%
\end{table}%
\subsection{Layout Area}
\begin{figure*}
        \centering 
       \begin{subfigure}[!t]{0.5\textwidth}
                \centering
                \includegraphics[width=\textwidth]{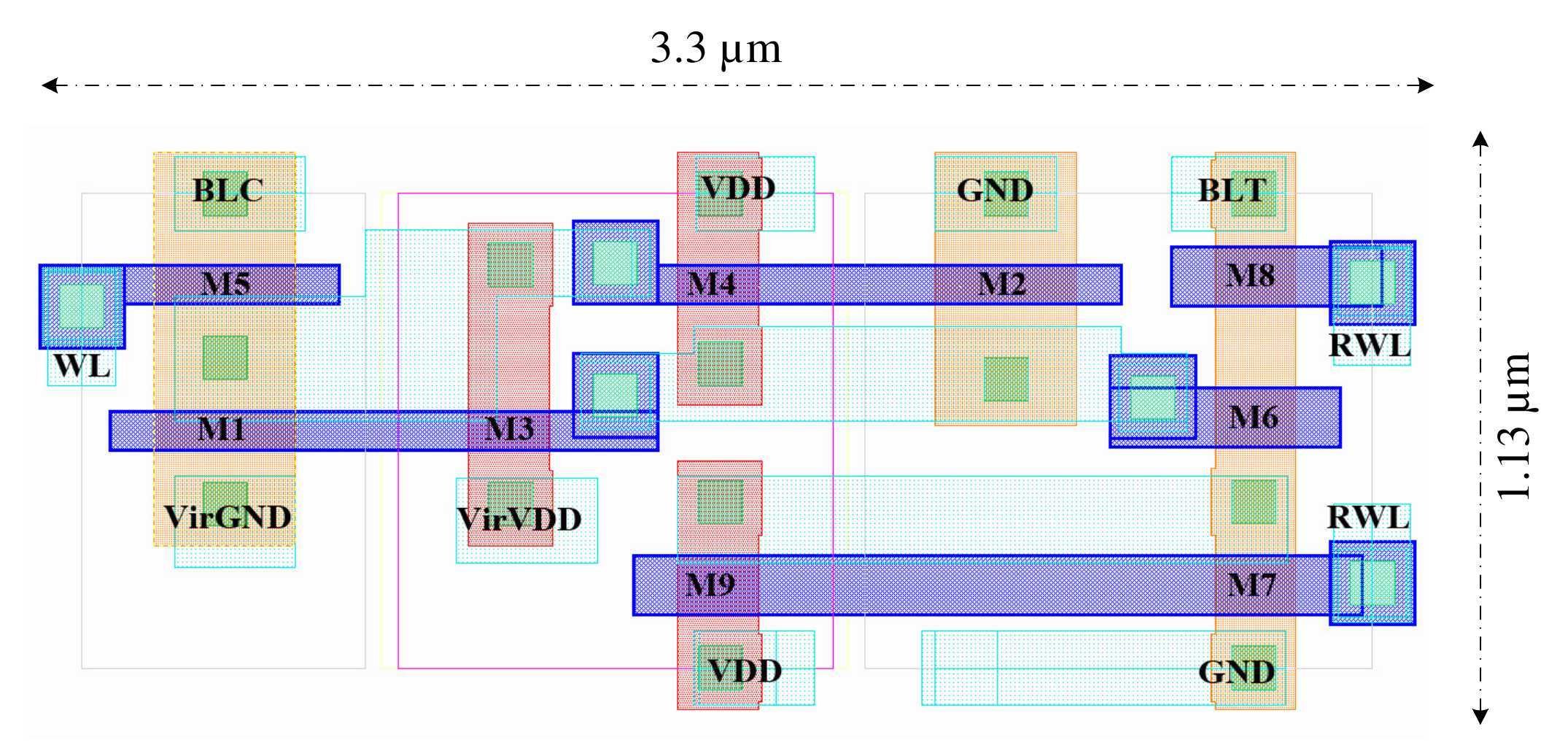}
                \caption{WRE9T SRAM cell}
                \label{Layout_WRE9T}
        \end{subfigure}%
        \begin{subfigure}[!t]{0.36\textwidth}
                \centering
                \includegraphics[width=\textwidth]{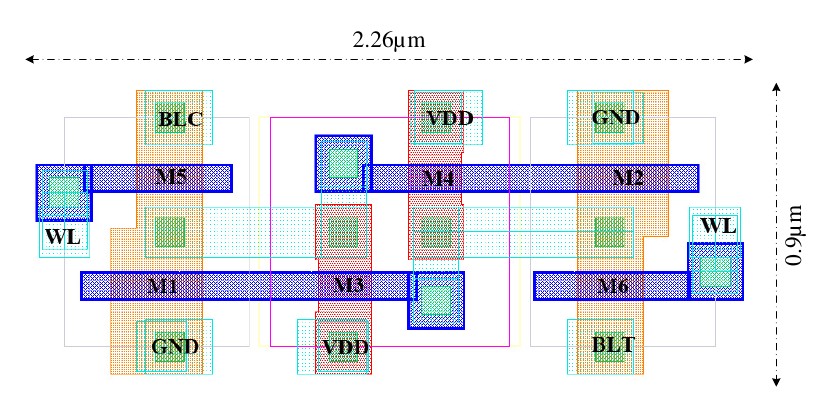}
                \caption{Conventional 6T SRAM cell}
                \label{Layout_6T}
        \end{subfigure}
        \caption{Layout of the proposed and conventional 6T SRAM cells in 90nm industrial CMOS technology.}\label{Layout_fig}
\end{figure*}

Fig. \ref{Layout_WRE9T} shows the layout of the proposed cell and Fig. \ref{Layout_6T} shows the layout of conventional 6T SRAM cell in a 90nm industrial CMOS technology. Table \ref{Layout_Areas} shows area of compared SRAM cells in this technology. As seen, conventional design has the minimum area, whereas ST-2 cell occupies larger area in the silicon.\\
Area of proposed WRE9T cell is larger than 6T by 83\%. However, simulation results for I.A.6T shows that, in the same consumption of silicon area for WRE9T and I.A.6T, $VDD_{min}$, read SNM, $I_{on}/I_{off}$ for read access path, are better for proposed design. Only hold SNM of I.A.6T is better than WRE9T cell. Therefore, conventional 6T SRAM cell with the same area cannot have capabilities similar to WRE9T. However, there is a drawback for WRE9T in physical implementation compared with conventional 6T SRAM cell; layout of WRE9T cell is not fully symmetric, and will be subject to systematic differences in processing variation.\\
\subsection{A 256kb SRAM}
We designed a 256kb SRAM using proposed WRE9T, and conventional 6T SRAM cells. As mentioned in Section \ref{sim:IonOFF}, worst case value of ${I_{on}}/{I_{off}}$ for WRE9T and 6T cells are 1080, and 78, respectively. Thus, we put 1k cells in each column of the proposed SRAM block and 64 cells in each column of SRAM block using 6T cells. Table \ref{256k_table} lists write and read power and delays for both of these SRAMs. We have mentioned the results assuming that each SRAM is operating at its minimum possible supply voltage. In this table, write delay is the time between activation of decoder until voltages of internal nodes of selected cell become equal. Read delay is the time between activation of decoder until 50mV voltage difference between two bit-lines occurs \cite{LP10T_Hasan} (for WRE9T that there is one bit-line for read operation, this difference is between voltage of read bit-line and Vdd). Write power and write delay reported here are average of writing 1 and 0, and read power is average of reading 1 and 0.  However, since we assume that bit-line is pre-charged to Vdd, read delay is the delay for reading 0. \\
\begin{table}[b]
  \centering
  \caption{Operating Parameters Comparison for 256kb SRAM Using WRE9T and Conv. 6T Cells.}
    \begin{tabular}{ccc}
    \toprule
      Design    & WRE9T@350mV & 6T@725mV \\
    \midrule
    Read Power ($\mu$W) & 25.68 & 376.5 \\
    Write Power ($\mu$W) & 32.75 & 456.12 \\
    Read Delay (ns) & 13.205 & 1.29 \\
    Write Delay (ns) & 12.88 & 1.76 \\
    Area (mm$^2$) &   1.18    &  0.86 \\
    \bottomrule
    \end{tabular}%
  \label{256k_table}%
\end{table}%
\begin{table}[t]
  \centering
  \caption{Overall design metric comparison.}
    \begin{tabular}{c c }
    \toprule
    Design metrics & The best  \\
    \midrule
    $I_{on}/I_{off}$ & WRE9T \\
    RSNM  						&  WRE9T  \\
    HSNM  						&  ST-2   \\
    WNM  			 			&  WRE9T  \\
    $VDD_{min}$ 			&  WRE9T  \\
    Leakage Power @$VDD_{min}$  & WRE8T  \\ 
   Cell Area                    &   6T       \\
    \bottomrule
    \end{tabular}%
  \label{Overall_Comp}%
\end{table}%
There are several papers such as \cite{Khellah,Zhao_GLSVLSI,zhai_JSSC} that discussed about single ended and differential read operations in SRAM cells, and their differences in different characteristics especially read delay. These comparisons and statements can be applied for all single ended and differential SRAM cells, and also to our single ended and conventional differential SRAM cells. However in this paper, we mostly focus on characterizing and comparing single SRAM cells. Thus, in the measurements of read delay, we did not consider the effect of sense amplifiers.\\
As seen, proposed design consumes lower power consumption for both read and write operations. Write and read circuitries of each column of SRAM block are shared among more cells when using new 9T cells compared to conventional 6T cells. This results in reducing area overhead of 256kb SRAM based on new 9T SRAM cells from 83\% (for comparing single SRAM cells) to 37\% compared to block design with conventional 6T.\\
In this section, several electrical design metrics were considered and the SRAM cells were compared and the best designs for each criterion were determined as listed in Table \ref{Overall_Comp}. Our proposed design WRE9T is the best in four metrics. Thus, we conclude this design can be the best choice to operate at low voltages.

\section{Conclusion}
\label{sec:conc}
In this paper a new 9T SRAM cell was presented. This SRAM cell improves write-ability and read stability at the same time. In this SRAM cell, $I_{on}/I_{off}$ of read access path is increased considerably that lets to integrate more cells in the same column of SRAM array. This property allows to share write and read peripheries that can directly translate to saving area and power. Proposed design decreases minimum possible operating voltage of SRAM by 375mV over conventional 6T and by 130mV over Iso Area 6T (I.A.6T) SRAM cell. Area of the proposed cell is larger than 6T cell by 83\%. This overhead is reduced to only 37\% by the potential of sharing write and read circuitry of each column in a 256kb SRAM.


%

\ifCLASSOPTIONcaptionsoff
  \newpage
\fi


%
\bibliographystyle{IEEEtran}
\bibliography{IEEEabrv,P9T1}

%
\begin{IEEEbiography}[{\includegraphics[width=1.1 in, height=3.3 in, clip, keepaspectratio]{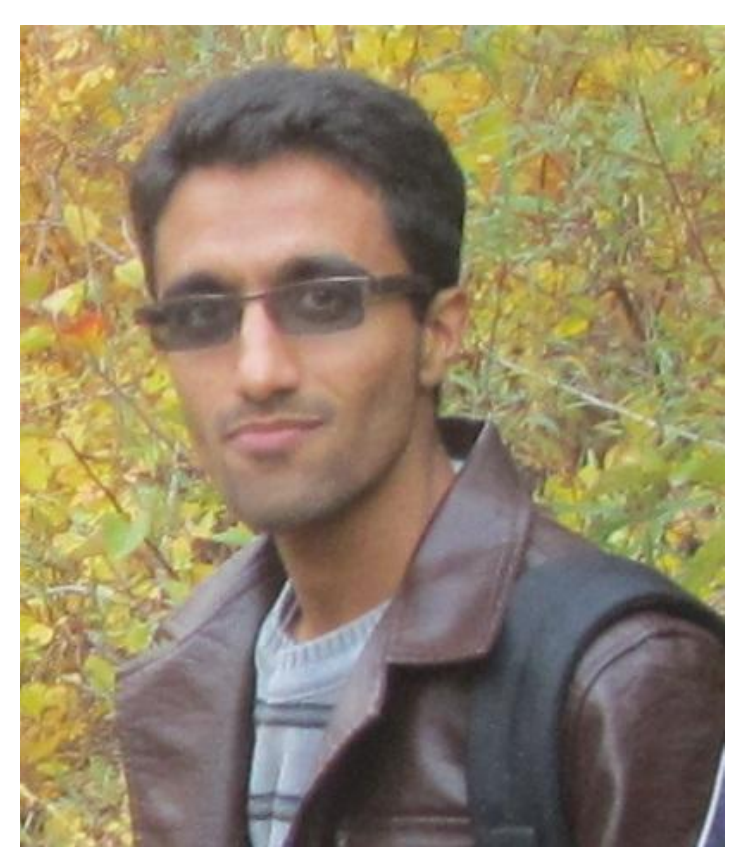}}] {Ghasem Pasandi}
(S’13) received the B.Sc, and M.Sc degrees in electrical and electronics engineering from the University of Tehran, Tehran, Iran in 2011, and 2014, respectively. He has published several conference and journal papers. His research interests  include Low Power and Ultra Low Power Static Random Access Memory (SRAM) design, Low Power and Energy Efficient logic design, reliable and fault tolerant static and dynamic CMOS circuit and system design, VLSI signal processing, and VLSI implementation of digital systems.
\end{IEEEbiography}

\begin{IEEEbiography}[{\includegraphics[width=1.1 in, height=3.9 in, clip, keepaspectratio]{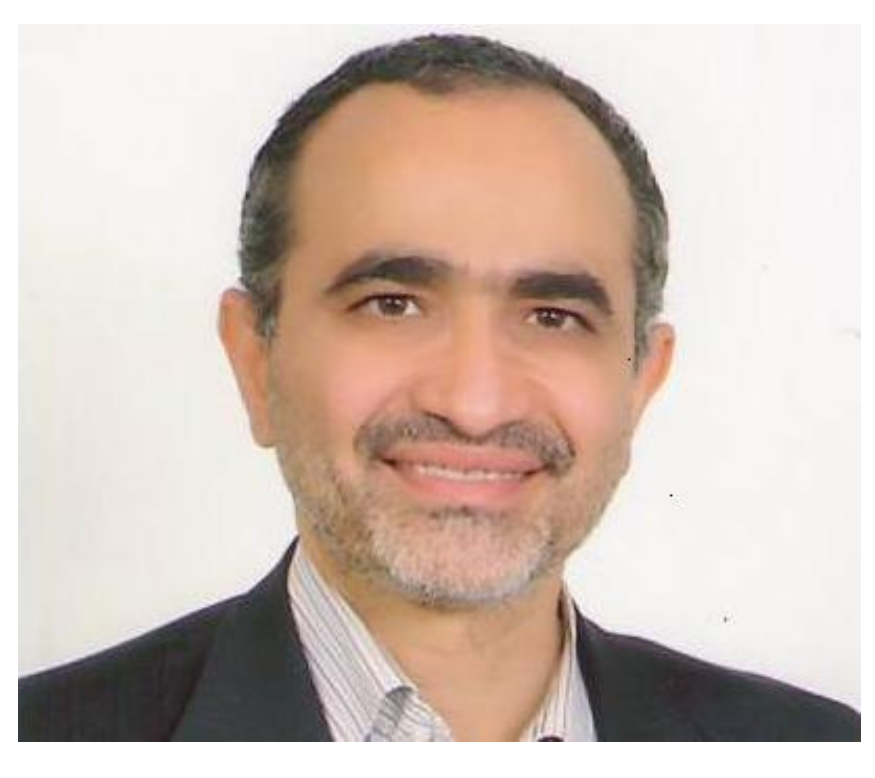}}]  {Sied Mehdi Fakhraie}
received his M.Sc. degree in electronics from the University of Tehran, Tehran, Iran, in 1989 and the Ph.D. degree in electrical and computer engineering from the University of Toronto, Toronto, ON, Canada in 1995. Since 1995, he has been with School of Electrical and Computer Engineering, University of Tehran, where he is now a Professor. He is Director of Silicon Intelligence and VLSI Signal Processing Laboratory. 
During the summers of 1998, 1999, and 2000, he was a visiting professor at the University of Toronto, where he continued his work on efficient implementation of artificial neural networks. He has also published more than 230 reviewed conference and journal papers. He has worked on many industrial IC design projects including design of network processors and home gateway access devices, DSL modems, pagers, and digital signal processors for personal and mobile communication devices. His research interests include system design and ASIC implementation of integrated systems, novel techniques for high-speed digital circuit design, and system-integration and efficient VLSI implementation of intelligent systems.
\end{IEEEbiography}




\end{document}